\documentstyle[12pt]{article}
\renewcommand\caption{}

\title{On Chiral Symmetry Breaking in a Constant Magnetic Field in Higher Dimension
}

\author{{\sl E. V. Gorbar} \\
{\sl{Instituto de Fisica Teorica, 01405-900 Sao Paulo, Brazil} \dag}
}

\date{}

\begin{document}

\maketitle

\vfill

\begin{abstract} 

Chiral symmetry breaking in the Nambu--Jona-Lasinio model in a constant magnetic field is studied in
spacetimes of dimension $D > 4$. It is shown that a constant magnetic field
can be characterized by [(D-1)/2] parameters. For the maximal number of nonzero field parameters,
we show that there is an
effective reduction of the spacetime dimension for fermions in the infrared region
D $\to$ 1 + 1 for even-dimensional spacetimes and D $\to$ 0 + 1 for odd-dimensional spacetimes.
Explicit solutions of the gap equation confirm our conclusions.

\end{abstract}
PACS 11.10.Kk, 11.30.Qc, 11.30.Rd \\
\dag On leave of absence from Bogolyubov Institute for Theoretical Physics, 252143,
Kiev, Ukraine

\vspace{2mm}

\vfill
\eject

\newpage

\newpage
\section{Introduction}

\vspace{1cm}

It was discovered recently \cite{PRL, GShM} that a constant magnetic field in 3+1 and 2+1 dimensions is a
strong catalyst of
dynamical chiral symmetry breaking leading to the generation of a fermion dynamical mass even at the weakest
attractive interaction between fermions. The essence of this effect is the dimensional reduction
of spacetime for fermions in the infrared region, which is $3 + 1 \to 1 + 1$ for $D = 3 + 1$ and
$2 + 1 \to 0 + 1$ for $D = 2 +1$. The dimensional reduction can be understood as follows.
The motion of charged fermions along the direction of magnetic field is free, therefore, the spectrum is continuos.
On the other hand, the motion in directions perpendicular to the magnetic field is restricted and
the spectrum is discrete
(fermions fill the Landau levels). Thus, the dynamics of fermions in a constant magnetic field
effectively corresponds in the infrared region
to the dynamics of fermions in (1 + 1)- and (0 + 1)-dimensional spacetimes in the cases of (3+1)-
and (2+1)-dimensional spacetimes, respectively.
(In this paper we consider chiral symmetry breaking in flat spacetimes of higher dimension
$D > 4$ with trivial topology (if topology of spacetime is not trivial, then there can be an additional reduction of
the spacetime dimension \cite{OCONNOR,FGI}).
To study this problem, we are motivated in addition to purely academic interest also by recent activity in
studing models with extra dimensions \cite{Arkani, Ran} and the availability of string solutions with
constant magnetic field (see, e.g., \cite{Tse}).

In Sect.2 we specify the Nambu--Jona-Lasinio (NJL) model \cite{NJL} in $D > 4$ and discuss the number of
parameters which characterize a constant magnetic field in $D > 4$.
We show in Sect.3 that, for the maximal number of field parameters, the effective
reduction of the spacetime dimension for fermions in the infrared region is D $\to$ 1 + 1 for even-dimensional
spacetimes and D $\to$ 0 + 1 for odd-dimensional spacetimes. We find the corresponding solutions
of the gap equation in the NJL model. Our conclusions are given in Sect.4.

\vspace{1cm}

\section{ The NJL model in a constant magnetic field }

\vspace{1cm}

To study dynamical chiral symmetry breaking in a constant magnetic field, we first need to classify
constant magnetic fields in $D > 4$, i.e., to define the number of independent parameters which specify a
constant magnetic field \cite{Gav}. Mathematically, the problem is the following:
A constant electromagnetic field is completely characterized by the field strength $F_{\mu\nu}$. Elements $F_{0\nu}$ and
$F_{\mu0}$ characterize the electric field. Elements $F_{ij}$, where i and j take values 1,..., n (D = n + 1),
define a constant magnetic field. By using orthogonal rotations, one can set some elements of
$F_{ij}$ to zero. It is a well-known fact of linear algebra that the number of independent parameters, which
define an arbitrary $F_{ij}$ up to orthogonal rotations, is $[\frac{n}{2}]$.

Let us present a simple inductive proof of this fact. Since $F_{ij}$ is antisymmetric, its diagonal elements
are zero. Obviously, $F_{ij}$ is characterized in general by $\frac{n(n-1)}{2}$ parameters, which we choose to be, e.g.,
the elements above the diagonal.
On the other hand, the orthogonal group in n-dimensional space has also $\frac{n(n-1)}{2}$ independent parameters because
there are
$\frac{n(n-1)}{2}$ independent rotations in n-dimensional space. Does it mean that we can set to zero all elements of
$F_{ij}$ by using appropriate orthogonal rotations? Of course, not. Explicit calculations show that an orthogonal
rotation in the plane $mn$ leaves unchanged the $F_{mn}$ element. Another fact is that if we perform a rotation in the
plane $kl$, where $kl$ takes value on the antidiagonal, then it leaves also all other antidiagonal elements unchanged.
It can be shown inductively for any $n \ge 3$ that one can set all elements of $F_{ij}$ to zero (except the elements
on the antidiagonal) by using orthogonal
rotations in planes $pq$, where $pq$ take all values except those on the antidiagonal. Since the remaining rotations
in planes $kl$, where $kl$ take values on the antidiagonal, do not change antidiagonal elements, the number of independent elements
of $F_{ij}$ is exactly the number of elements on the antidiagonal, which is obviously $[\frac{n}{2}]$.

We can assume without loss of generality that the magnetic part of the field strength $F_{\mu\nu}$ in a convenient
reference frame is given by
\begin{equation}
F_{ij} = \sum_{k=1}^{[\frac{n}{2}]} H_k(\delta_{i}^{k}\delta_{j}^{n+1-k} - \delta_{j}^{k}\delta_{i}^{n+1-k})
\end{equation}
and the corresponding vector potential is
\begin{equation}
A_i = - H_i x_{n+1-i}.
\end{equation}

Let us now consider dynamical chiral symmetry breaking in the NJL
model in a constant magnetic field in $D > 4$. We first discuss
what we mean by chiral symmetry in spacetimes of arbitrary dimension.
As well known, the notion of chiral symmetry is connected with properties of representations of the Clifford
algebra (for a very clear discussion see, e.g., \cite{West}).
The Clifford algebra for spacetimes of even dimension has only one
complex irreducible representation in the $2^{D/2}$-dimensional
spinor space. These spinors are reducible with respect to the even
subalgebra (generated by products of an even number of Dirac matrices)
and split in a pair of
$2^{D/2-1}$-component irreducible Weyl spinors ($\gamma_{D} = \gamma_0 ... \gamma_{D-1}$ is
an analog of the $\gamma_5$ matrix in D-dimensional spacetime and $\frac{1 \pm \gamma_{D}}{2}$ are the corresponding
chiral projectors). In odd-dimensional spacetimes, there are two different representations of the Clifford algebra (they
differ by the sign of the $\gamma$-matrices) and chiral symmetry is not defined because $\gamma_{D}$ is proportional
to the unity. In order to define an analog of chiral symmetry in odd-dimensional spacetimes, it is the usual practice
to assume that
fermion fields are in a reducible representation of the Clifford algebra so that we can define an analog of chiral
symmetry (for an explicit example in
(2 + 1)-dimensional spacetime see, e.g., \cite{App}). In what follows
we understand chiral symmetry in odd-dimensional spacetimes in this sense.

For our aims it is enough to consider the following generalization of the NJL model with $U_{\rm L}(1)\times
U_{\rm R}(1)$ chiral symmetry to $D > 4$:
\begin{eqnarray}
{\cal{L}}=\bar{\psi} i\gamma^\mu D_\mu \psi+
{G\over 2}\bigl[(\bar\psi\psi)^2+
(\bar\psi i\gamma_D\psi)^2\bigr]~,
\end{eqnarray}
where $D_\mu =  \partial_{\mu} + ie A_{\mu}$ is the covariant derivative and fermion fields carry an additional 'flavor'
index $i = 1, \ldots, N$.

\vspace{1cm}

\section{Dynamical chiral symmetry breaking}

\vspace{1cm}

By introducing auxiliary fields, we can rewrite Lagrangian (3) in the following way:
\begin{eqnarray}
{\cal{L}}= \bar{\psi} i\gamma^\mu D_\mu\psi-
\bar\psi(\sigma+i\gamma_D\pi)\psi-{1\over 2G}
(\sigma^2+\pi^2)~.
\end{eqnarray}
Indeed, the Euler-Lagrange equations for the auxiliary fields $\sigma$ and $\pi$ are
\begin{eqnarray}
\sigma=-G(\bar\psi\psi)~, \quad
\pi=-G(\bar\psi i\gamma_D\psi)~,
\end{eqnarray}
and Lagrangian (4) gives Lagrangian (3) if we use the equation of motion (5).

By integrating over fermions, we get the following effective action for the composite fields:
\begin{equation}
\Gamma(\sigma,\pi)=
-i{\rm Tr}\,{\rm Ln}\bigl[i \hat D -
(\sigma+i\gamma_D\pi)\bigr]
-{1\over 2G} \int
d^Dx (\sigma^2+\pi^2)~,
\end{equation}
where $\hat D = \gamma^\mu D_\mu$.
As usual in calculation of the effective potential, it is enough to set $\sigma=const$ and
$\pi=const$. Since the effective potential $V$ depends only on the
$U_{\rm L}(1)\times U_{\rm R}(1)$-invariant $\rho^2=\sigma^2+\pi^2$,
it is sufficient to consider a configuration with $\pi=0$ and $\sigma=const$. 
Since
\begin{eqnarray}
{\rm Det}(i\hat D-\sigma)={\rm Det}(\gamma_D(i\hat D-\sigma)\gamma_D)=
{\rm Det}(-i\hat D-\sigma)~,
\end{eqnarray}
we find that
\begin{eqnarray}
{\rm Tr}\,{\rm Ln}(i\hat D-\sigma)
=-{i\over 2}{\rm Tr}
\bigl[\,{\rm Ln}(i\hat D-\sigma)+
{\rm Ln}(-i\hat D-\sigma)\bigr] 
=-{i\over 2}{\rm Tr}\,{\rm Ln}(\hat D^2+\sigma^2)\, . 
\end{eqnarray}
By using the method of proper time, we have
\begin{eqnarray}
-{i\over 2}{\rm Tr}\,{\rm Ln}(\hat D^2+\sigma^2)
={i\over 2} \int d^Dx \int^\infty_0 {ds\over s} {\rm tr}
\langle x|e^{-is(\hat D^2+\sigma^2)}|x\rangle,
\end{eqnarray}
where
\begin{eqnarray}
\hat D^2=D_\mu D^\mu-{ie\over 2}
\gamma^\mu\gamma^\nu F_{\mu\nu}.
\end{eqnarray}
For $A_{\mu}$ given by Eq.(2), it is obvious that the problem of calculation of
the matrix element $\langle x|e^{-is(\hat D^2+\sigma^2)}|x\rangle$ is reduced
to the calculation of the corresponding matrix element for every
$H_k$, i.e. for $x_k$ and $x_{n+1-k}$ components. By using \cite{Schwi}, we obtain
the following effective potential:
\begin{eqnarray}
V(\rho)= \frac{\rho^2}{2G} +
\frac{2^{[\frac{D+1}{2}]} N}{2(4\pi)^{d/2}} \int^\infty_{1/\Lambda^2}
\frac{ds}{s^{D/2 - [(D-1)/2]+1}} e^{-s\rho^2} \prod_{k=1}^{[(D-1)/2]} eH_k\coth (eH_ks)~,
\end{eqnarray}
where $\Lambda$ is a ultraviolet cutoff.
The gap equation $\frac{dV}{d\rho} = 0$ takes the form
\begin{equation}
\frac{1}{G} = \frac{2^{[\frac{D+1}{2}]} N}{(4\pi)^{D/2}} \int^\infty_{1/\Lambda^2}
\frac{ds}{s^{D/2-[(D-1)/2]}} e^{-s\rho^2} \prod_{k=1}^{[(D-1)/2]} eH_k \coth(eH_ks)~.
\end{equation}

If the magnetic field is absent, then the right-hand side of the gap equation is
\begin{equation}
\frac{2^{[\frac{D+1}{2}]} N}{(4\pi)^{D/2}} \int^\infty_{1/\Lambda^2}
\frac{ds}{s^{D/2}} e^{-s\rho^2}~,
\end{equation}
where the integrand is exactly the heat kernel of the Dirac operator squared in D-dimensional spacetime.
Since $\coth x \to 1$ as $x \to \infty$, it follows from Eq.(12) that every independent parameter
of magnetic field $H_k$, which is not equal to zero, effectively
reduces the spacetime dimension by 2 units in the infrared region (for $ s \to \infty$ only the lowest part of the
spectrum of the Dirac operator squared gives contribution). Consequently, for the maximal number of
field parameters $[\frac{D-1}{2}]$, we obtain that the effective reduction of the spacetime dimension in
the infrared region for fermions is
$D \to 1 + 1$ for even-dimensional spacetimes and $D \to 0 + 1$ for odd-dimensional spacetimes. Thus, we expect that
for the maximal number of field parameters the critical coupling constant
is zero for even-dimensional spacetimes and the gap analytically depends on coupling constant in odd-dimensional
spacetimes, which are the
characteric features of solutions of the gap equation in two- and one-dimensional spacetimes, respectively
(see \cite{PRL, GShM}).

To analyze the gap equation, we assume for simplicity that all $H_k$ are equal, i.e. $H_1=H_2=\ldots=H_{[(D-1)/2]}=H$.
Since we are mainly interested only in qualitative results, we split the interval of integration in two parts from
$\frac{1}{\Lambda^2}$ to $\frac{1}{eH}$ and from $\frac{1}{eH}$ to $\infty$ and approximate $\coth x$ by $1/x$ on
the first interval and by 1 on the second (we assume also that $\rho^2 \ll eH$ and approximate $e^{-s \rho^2}$ by 1
on the first interval). One can check that this approximation for D=3 and D=4
gives the same result for the gap (the same dependence on $eH$, $\Lambda^2$, and G) as
the exact result \cite{PRL, GShM} up to a numerical constant of order O(1).  For even D, we obtain the following
gap equation:
\begin{equation}
\frac{(2\pi)^{D/2}}{G N \Lambda^{D-2}} = \frac{1 - (eH/\Lambda^2)^{D/2 - 1}}{D/2 - 1} +
(eH/\Lambda^2)^{D/2 - 1} \int_{\rho^2/eH}^{\infty} \frac{ds}{s} e^{-s}.
\end{equation}
By using \cite{GR}
\begin{eqnarray*}
\Gamma (\alpha, x) = \int_x^{\infty} e^{-t} t^{\alpha - 1}
\end{eqnarray*}
and an expansion of the incomplete Gamma-function for small $x$
\begin{equation}
\Gamma (0, x) = -C - \ln x - \sum_{k=1}^{\infty} \frac{(-x)^k}{k \cdot k!},
\end{equation}
where C is the Euler constant, we obtain the solution
\begin{equation}
\rho^2 = eH exp^{-\frac{(2\pi)^{\frac{D}{2}}(1-g)}{GN(eH)^{\frac{D}{2}-1}}},
\end{equation}
where $g = \frac{GN\Lambda^{D-2}}{(D/2-1)(2\pi)^{\frac{D}{2}}}$.
For odd D, the gap equation is
\begin{equation}
\frac{(2\pi)^{D/2}}{G N \sqrt{2}} = \frac{\Lambda^{D-2}(1 - (eH/\Lambda^2)^{D/2 - 1})}{D/2 - 1} +
\frac{(eH)^{(D-1)/2}}{\rho} \int_{\rho^2/eH}^{\infty} \frac{ds}{s^{1/2}} e^{-s}.
\end{equation}
By using \cite{GR}
\begin{equation}
\Gamma (\alpha, x) = \Gamma(\alpha) - \sum_{k=0}^{\infty} \frac{(-1)^k x^{\alpha+k}}{k! (\alpha+k)} \,\,\,\,\,\,\,\,
(\alpha \ne 0, -1, -2, \ldots),
\end{equation}
we obtain the solution
\begin{equation}
\rho = \frac{(eH)^{\frac{D-1}{2}} G N}{(2\pi)^{\frac{D-1}{2}}}.
\end{equation}
(Note that $\rho^2 \ll eH$ for sufficiently small G and the approximation of $e^{-s\rho^2}$ by 1 on the interval
$[\frac{1}{\Lambda^2}, \frac{1}{eH}]$ is consistent.)
Thus, for the maximal number of field parameters,
the critical coupling constant is zero in even-dimensional spacetimes and the gap has an essential singularity
at zero value of coupling constant. In odd-dimensional spacetimes
the gap depends analytically on coupling constant. These results confirm our
conclusions about the effective reduction of the spacetime dimension for fermions in the infrared region because
our solutions are characteric for solutions of the gap equation in (1+1)- and (0+1)-dimensional
spacetimes, respectively \cite {PRL, GShM}.

Let us consider briefly the case where only $m < [\frac{D-1}{2}]$ field parameters are not
equal to zero. As follows from the gap equation (12) the critical coupling constant is not equal to zero in
this case. The most interesting for us is the dependence of the critical coupling constant on the number of
field parameters and the value of magnetic field.
For simplicity we again assume that all $H_k$ are equal, i.e. $H_1=H_2=\ldots=H_{m}=H$. By approximating
$\coth x$ by 1 in the interval $[\frac{1}{eH}, \infty]$ and $1/x + x/3$ in the interval 
$[\frac{1}{\Lambda^2}, \frac{1}{eH}]$, we find that the critical coupling constant is equal to
(due to the approximations made, the coefficients near terms
$(\frac{eH}{\Lambda^2})^{\frac{3}{2}}$, $(\frac{eH}{\Lambda^2})^2$, and
$(\frac{eH}{\Lambda^2})^2 \ln\frac{\Lambda^2}{eH}$ are actually defined up to a constant of order 1)
\begin{equation}
g_{cr} = \frac{1}{1 + 2(\frac{eH}{\Lambda^2})^{\frac{3}{2}}}
\end{equation}
for D=5,
\begin{equation} 
g_{cr} = \frac{1}{1 + (\frac{eH}{\Lambda^2})^2 + \frac{2}{3}(\frac{eH}{\Lambda^2})^2 \ln\frac{\Lambda^2}{eH}}
\end{equation}
for D=6, and
\begin{equation} 
g_{cr} = \frac{1}{1 + \frac{m(\frac{D}{2}-1)}{3(\frac{D}{2}-3)}(\frac{eH}{\Lambda^2})^2}
\end{equation}
for $D > 6$. For D = 5 and D = 6, a constant magnetic field is characterized by only two independent field parameters,
therefore, $m$ can be equal only to 1 in the case under consideration. As follows from Eq.(22)
for $D > 6$ the more the number of field parameteres $m$, the less the critical coupling constant
in agreement with expectations. Eqs.(20)-(22) imply that the critical coupling constant
is very close to 1 in the realistic case $eH \ll \Lambda^2$.

\vspace{1cm}

\section{Conclusions}

\vspace{1cm}

We considered chiral symmetry breaking in the NJL model in a constant magnetic field in $D > 4$.
We showed that for the maximal number of field parameters the effective
reduction of the spacetime dimension for fermions in the infrared region is D $\to$ 1 + 1 for even-dimensional
spacetimes and D $\to$ 0 + 1 for odd-dimensional spacetimes. We studied the gap equation of the
NJL model and found that for the maximal number of
field parameters the gap is analytic in coupling constant for odd-dimensional spacetimes and
the gap has an essential singularity at zero value of coupling constant for even-dimensional spacetimes, which are
characteric features of solutions of the gap equation in 0 + 1 and 1 + 1 dimension, respectively, that confirms
the dimensional reduction. We would like to note also that our results can be relevant in the context of
certain string solutions \cite{Tse}
in the low-energy domain, where we can have a constant magnetic field in spacetimes with dimension $D > 4$.

The author thanks I.L. Buchbinder, S.P. Gavrilov, D.M. Gitman, A.A. Natale, and F. Toppan for useful discussions and
valuable remarks. I am grateful to V.P. Gusynin for reading the manuscript and useful suggestions.
This work was supported in part by FAPESP grant No. 98/06452-9.

\end{document}